\documentclass[a4paper,11pt]{article}
\newcommand{\pp}[1]{\frac{\partial}{\partial#1}}
\newcommand{\p}{\partial}

\def\D{\mathord{\mkern7mu\mathaccent"7020{\mkern-7mu\partial}}}
\def\PPh#1{\setbox0\hbox{$#1\rm I$}\vcenter{\ialign{$#1\rm##$\cr
I\cr\noalign{\nointerlineskip\vskip-0.541\ht0}P\cr}}}
\def\Ph{{\mathord{\mathpalette\PPh{}}}}
\newcommand{\Php}{\Ph^{\prime}}
\newcommand{\Dp}{\D^{\, \prime \!}}
\newcommand{\ei}{e^{-1}}
\newcommand{\beqn}{\begin{eqnarray}}
\newcommand{\eeqn}{\end{eqnarray}}

\newcommand{\pa}{\frac{\partial}{\partial \alpha}}
\newcommand{\pat}{\frac{\partial^2}{\partial \alpha^2}}
\newcommand{\dpa}{\frac{d}{d \alpha}}
\newcommand{\dpat}{\frac{d^2}{d \alpha^2}}
\newcommand{\al}{\alpha}

\newcommand{\spp}{\cal I^+}
\newcommand{\sm}{\cal I^-}

\newcommand{\pab}[3]{#1_{\mskip-4mu{\scriptstyle #2}^{\mskip4mu\scriptstyle
#3}}}

\newtheorem{proposition}{Proposition}[section]

\newtheorem{definition}[proposition]{Definition}
\newtheorem{example}[proposition]{Example}

\let\Definition=\definition
\renewcommand{\definition}{\Definition \rm}
\let\Example=\example
\renewcommand{\example}{\Example \rm}

\title{On the differentiability conditions at spacelike infinity}
\author{Magnus Herberthson\thanks{Link\"{o}ping University, Department of
Mathematics,
S-581 83 Link\"{o}ping, Sweden}}

\begin{document}
\maketitle
\begin{abstract}
We consider space-times which are asymptotically flat at spacelike
infinity, ${i^0}$. It is well known that, in general, one cannot
have a smooth differentiable structure at ${i^0}$, but have to use direction
dependent structures. Instead of the oftenly used $C^{>1}$-differentiabel
structure, we suggest a weaker differential structure, a $C^{1^+}$ structure.
The reason for this is that we have not seen any completions of the
Schwarzschild space-time which is $C^{>1}$ in both spacelike and null
directions at ${i^0}$. In a $C^{1^+}$ structure all directions can be treated
equal, at the expense of logarithmic singularities at ${i^0}$. We show that,
in general,
the relevant part of the curvature tensor, the Weyl part, is free from these
singularities, and that the (rescaled) Weyl tensor has a certain symmetry
property.
\end{abstract}
\newpage

\section{Introduction}
When studying asymptotic properties of space-times, one often has to use
what is called direction dependent structures. In \cite{asht1}, Ashtekar
and Hansen
examines space-times that are asymptotically flat at null and spacelike (or
spatial) infinity. Using conformal completions, spacelike infinity is
represented by  a point, $i^0$, which is the vertex of $\spp$ and $\sm$.

In general, the differential structure at $i^0$ cannot be smooth, and in
\cite{asht1}
a so called $C^{>1}$ differential structure is used. Roughly speaking, the
(conformally rescaled) metric $g_{ab}$ is well defined at $i^0$, the first
derivatives of
the metric are direction dependent at $i^0$, and the curvature components
diverge
like $1/r$ as one approaches $i^0$ (in spacelike directions). Here $r$ is
some suitable distance function with $r(i^0)=0$. It is worth noting that in
the definition
of a $C^{>1}$ differential structure, as given in \cite{asht1}, one examines
direction dependent quantities by their limits at $i^0$ along spacelike
directions only. As we will see, in the completions of the ``obviously''
asymptotically flat
Schwarzschild space-time which are given by e.g.
\cite{asht1},\cite{chrusciel}, one has
rescaled metrics which are $C^{>1}$ at $i^0$ in this sense, but where where the
metric is only $C^0$ on $\spp$ and $\sm$. In the following we assume that the
reader is familiar with the concept of $C^{>n}$ differentiable structures
and functions as defined in \cite{asht1} or \cite{wald}, although we
for completeness repeat some basic definitions in section
\ref{udiffcond}.

There are examples of other conformal rescalings of the Schwarzschild
space-time,
\cite{chrusciel},\cite{schmidtstew}, where the rescaled metric is smooth at
$\spp$ and $\sm$, but
where the derivatives of the metric do not have direction dependent limits
at $i^0$ but rather diverges logarithmically. At first this may seem as a
serious
defect, since these logarithmic singularities can be expected to show up in the
conformally rescaled Ricci and Riemann tensors. This is certainly the case,
but as we will show below,
for the physically relevant part, namely the Weyl tensor, these singularities
disappear.

Below we start by comparing the two different situations mentioned
above; this will serve as a motivation for the new differential
structure that we then introduce. Finally we show that an expected
singularity for the Weyl tensor is absent, and we also discuss some of
the consequencies of this.

\section{Two completions of Schwarzschild space-time}
\label{utwocomp}
As a motivation for the differentiability conditions introduced in
the following section, we compare two different completions of the
Schwarzschild
space-time. These completions can be found in for instance
\cite{asht1},\cite{chrusciel},\cite{schmidtstew}.
In the Schwarzschild space-time, with standard
coordinates $t,r,\theta
,\phi$ the metric is
\beqn
ds^2=(1-\frac{2m}{r})dt^2-\frac{1}{1-\frac{2m}{r}}dr^2-r^2d\Sigma^2,\quad
d\Sigma^2=d\theta^2+sin^2\theta\,d\phi^2.
\eeqn
A Schwarzschild space-time is of course considered as asymptotically
flat at
spacelike infinity, $i^0$, as well as at null infinity, ${\cal I}$.
Therefore,
any reasonable definition of asymptotic flatness in these regions
should clearly
admit this space-time.
For readers not familiar with the concept of asymptotic flatness
we refer to
\cite{asht1} and \cite{wald} for definitions and discussions, although
some definitions are included in the next section.

The first conformal rescaling we look at uses the so called Schmidt-Walker
coordinates,
\cite{asht1},\cite{chrusciel}. This gives us a ``standard''
completion of the
Schwarzschild space-time, in which the new, unphysical metric
$\widehat g_{ab}$ is
$C^{>0}$ at $i^0$ only for spacelike directions.
Here $\widehat g_{ab}=\Omega^2g_{ab}$, where $g_{ab}$ is the physical
metric and
$\Omega$ is the conformal factor.\\
Namely, let us
first define, for $r>2m$, the function $f$ by
\beqn
f(r)=r+2m\log(\frac{r}{2m}-1).
\eeqn
Put $r^\star=f(r)$ and define $\widehat v$ and $\widehat w$
implicitly by
\beqn
r^\star=\frac{1}{2}[f(\frac{1}{\widehat v})+f(\frac{1}{\widehat w})],
\quad
t=\frac{1}{2}[f(\frac{1}{\widehat v})-f(\frac{1}{\widehat w})].
\label{comp1}
\eeqn
With $\widehat\Omega=\widehat v \widehat w$, the rescaled metric
$\widehat g_{ab}=\widehat\Omega^2g_{ab}$
becomes
\beqn
d\widehat s^2=-\frac{1-2m/r}{(1-2m\widehat v)(1-2m \widehat w)}
d\widehat v d\widehat w-
\omega^2\Bigl(\frac{\widehat v+\widehat w}{2}\Bigr)^2 d \Sigma^2,
\eeqn
where $\omega=\omega(\widehat v,\widehat w)$.
Here $i^0$ is of course the point $\widehat v=\widehat w=0$.
$\widehat v=0$ and $\widehat w=0$
corresponds to $\spp$ and $\sm$ respectively. If we put
${\hat t}=\frac{{\hat v}-{\hat w}}{2}$ and ${\hat r}=
\frac{{\hat v}+{\hat w}}{2}$,
 $\omega$ can be brought to the form
(see \cite{chrusciel})
$$
\omega=1+m \widehat r(1-\frac{\widehat t^2}{\widehat
r^2})\log(1-\frac{\widehat t^2}{\widehat r^2})+
O(\widehat r^2).
$$
This means that $\omega$ takes a finite value (by continuity) on $\cal
I$
i.e., when $\frac{\hat t}{\hat r}=\pm 1$, but
that non-tangential derivatives do not exist there.
Therefore, on ${\cal I}$, $\widehat g_{ab}$ is
defined but not differentiable.
For spacelike directions approaching $i^0$, the situation is better.
The metric $\widehat g_{ab}$  is $C^{>0}$ at $i^0$; in particular
$\sqrt {\widehat \Omega} \widehat R_{abcd}$ has a
regular direction dependent limit for these directions.
Here $\widehat R_{abcd}$ is of course the Riemanntensor connected to
$\widehat g_{ab}$.

The second completion is similar, but
has
different properties. In this completion, taken from
\cite{schmidtstew}
(see also remark in \cite{chrusciel}),
the metric will not be $C^{>0}$ at $i^0$ in any direction. On the
other hand, it
will be smooth on ${\cal I}$. Also, the metric will have components
where the
derivatives diverges in a ``mild'' way, namely logarithmically with
respect to a suitable radial parameter.\\
This time, cf.\cite{schmidtstew}, we put
\beqn
r^\star=[f(\frac{1}{\widetilde v})+f(\frac{1}{\widetilde w})], \quad
t=[f(\frac{1}{\widetilde v})-f(\frac{1}{\widetilde w})], \nonumber \\
\widetilde t=\frac{\widetilde v-\widetilde w}{2}, \quad \widetilde
r=\frac{\widetilde v+ \widetilde w}{2},\quad \widetilde
\Omega=\widetilde v\widetilde w. \label{comp2}
\eeqn
The rescaled metric becomes
\beqn
d\widetilde s^2=-\frac{1-2m/r}{(1-2m\widetilde v)(1-2m \widetilde w)}
d\widetilde v d\widetilde w-
\Bigl(\frac{2\widetilde v \widetilde w r }{\widetilde v+\widetilde
w}\Bigr)^2 \widetilde r^2 d \Sigma^2.
\eeqn
As shown in \cite{schmidtstew}, $\frac{\tilde{v} \tilde w r }{\tilde
v+\tilde w}$ has the
following expansion in a neighbourhood of $i^0$;
\beqn
\frac{\widetilde v \widetilde w r }{\widetilde v+\widetilde
w}=1-2m\frac{\widetilde v \widetilde w }{\widetilde v+\widetilde
w}\log(2m(\widetilde v+\widetilde w))+O([(\widetilde v+\widetilde
w)\log(\widetilde v+\widetilde w)]^2).
\eeqn
This expression can be differentiated arbitrarily many times. In
particular,
$\frac{\tilde v \tilde w r }{\tilde v+\tilde w}r$ is shown to be
analytic on
$\spp$ and $\sm$. Thus in this completion, the metric is smooth on
$\spp$ and
$\sm$, but it is not $C^{>0}$ at $i^0$. In fact, the metric is
continuos at
$i^0$ and has derivatives that blow up logarithmically (i.e. like
$\log \widetilde r$
as $\widetilde r \to 0^+$) as one approaches $i^0$. Comparing with
definition
\ref{C0pdeff} below, we see that the metric is $C^{0^+}$ at $i^0$.

We will now discuss some properties of these completions. Let us
first recall
that the Weyl tensor ${C_{abc}}^d$ is conformally invariant, i.e.,
under a
rescaling of the metric $g_{ab} \to \widehat g_{ab}=\widehat\Omega^2
g_{ab}$, the
Weyl tensor transforms like ${C_{abc}}^d \to {\widehat
C_{abc}}^{\phantom {ddd} d}
={C_{abc}}^d$.
Let us also recall that, for a space-time which is asymptotically
flat at null
infinity, $\Omega^{-1}{C_{abc}}^d$ has a smooth limit at $\cal I$.
(For a proof of this, see \cite{pen2}.) In our first completion,
using the coordinates
given by
(\ref{comp1}), the rescaled metric $\widehat g_{ab}$ is not smooth on
$\cal I$, so
we can not say, a priori, anything about the limit of
$\widehat\Omega^{-1}{\widehat C_{abc}}^{\phantom {ddd}d}$
there. Of course, in the special case of a Schwarzschild space-time,
we can check
directly that $\widehat\Omega^{-1}{\widehat C_{abc}}^{\phantom
{ddd}d}$ happens to have a limit on $\cal I$.
On the other hand is $\widehat R_{abcd}$ singular on $\cal I$.

Similarly, for the second completion, using (\ref{comp2}) where the
metric is not
$C^{>0}$ in
spacelike directions at $i^0$, we expect that $\sqrt
{\widetilde\Omega} \widetilde R_{abcd}$
diverges logarithmically at $i^0$. Nevertheless, for the space-time
considered,
$\sqrt {\widetilde\Omega} \widetilde C_{abcd}$ will have direction
dependent limits at $i^0$.

Basically,
\begin{enumerate}
\item $\widehat g_{ab}$ is $C^{>0}$ in spacelike directions from
$i^0$, so that $\lim_{i^0} \sqrt {\widehat \Omega} {\widehat
C_{abc}}^{\phantom
{ddd}d}$ exists along spacelike directions.
\item $\widetilde g_{ab}$ is smooth on $\cal I$ so that
$\widetilde \Omega^{-1}{\widetilde C_{abc}}^{\phantom{ddd}d}$ exists
on $\cal I$.
\end{enumerate}
By the conformal invariance we have that
\beqn
\sqrt {\widetilde \Omega} {\widetilde C_{abc}}^{\phantom {ddd}d}=
\sqrt{\frac{\widetilde \Omega}{\widehat \Omega}}
\sqrt {\widehat \Omega} {\widehat C_{abc}}^{\phantom{ddd}d}=
\sqrt{\frac{\widetilde v}{\widehat v}
\frac{\widetilde w}{\widehat w}}
\sqrt {\widehat \Omega} {\widehat C_{abc}}^{\phantom{ddd}d}.
\eeqn
and
\beqn
\widehat \Omega^{-1}{\widehat C_{abc}}^{\phantom{ddd}d}=
\frac{\widetilde \Omega}{\widehat \Omega}
\widetilde \Omega^{-1}{\widetilde C_{abc}}^{\phantom{ddd}d}=
\frac{\widetilde v}{\widehat v}
\frac{\widetilde w}{\widehat w}
\widetilde \Omega^{-1}{\widetilde C_{abc}}^{\phantom{ddd}d}.
\eeqn
By examining $\frac{\widetilde v}{\widehat v}
\frac{\widetilde w}{\widehat w}$ it is not hard to see that this
expression has a (non-differentiable) non-zero limit as one appraoches
$\cal I$ or $i^0$. Thus, in addition to the limits above, we can also
conclude that
$\lim_{i^0} \sqrt {\widetilde \Omega} {\widetilde C_{abc}}^{\phantom {ddd}d}$
exists and that
$\widehat \Omega^{-1}{\widehat C_{abc}}^{\phantom{ddd}d}$ exists (by
continuity) on $\cal I$, although this was not expected from the
behaviour of the metrices $\widehat g_{ab}$ and $\widetilde g_{ab}$.
Thus, for the Weyl tensor (at least in this case),
one can allow the metric to be less regular
than expected, and still have `nice limits' at $i^0$.
This idea will now be developed.

\section{The differentiability conditions at $i^0$}
\label{udiffcond}
In this section we will discuss the differentiability conditions at
spacelike
infinity $i^0$, aiming at
emphasizing the (physical) gravitational field rather than the metric.
These changes will
be motivated by the previous example and because, to our knowledge,
there is no
given example of a conformal rescaling
of the Schwarzschild metric which makes the rescaled metric $C^{>0}$
in both
spacelike and null directions.

Since from now on we will deal mainly with conformally rescaled
space-times,
we adopt the convention that the rescaled metric (normally) will be
written
without hat/tilde. Thus we may have $g_{ab}=\Omega^2 \pab gp{ab}$,
where
$\Omega$ is the conformal factor and $\pab gp{ab}$ the physical
metric.
Accordingly, $R_{abcd},R_{ab}$ etc. will refer to quantities related
to the
unphysical metric $g_{ab}$ and the associated derivative operator.

We believe that the interest should be focused, not on the metric,
but rather
on the Weyl tensor $C_{abcd}$. The $C^{>0}$ condition at $i^0$ is
precisely what
is needed in order to deduce that $\sqrt \Omega R_{abcd}$ is regular
direction dependent at $i^0$. However, starting with a (physical)
vacuum
space-time, the physical Ricci tensor $\pab Rp{ab}$ is zero and all
information about the
space-time lies within the Weyl tensor. After the rescaling, the Weyl
tensor
remains the same, i.e., carries the same information (about the
physical
space-time) as before. In addition, due to the conformal
transformation,
we get, see \cite{wald}, a non-physical Ricci tensor $R_{ab}$.
We argue that the condition that
$\sqrt \Omega R_{ab}$ is regular direction dependent at $i^0$ may be
too
strong and perhaps of secondary interest. Instead we should try to
ensure
that $\sqrt \Omega C_{abcd}$ is regular direction dependent at $i^0$.
(Of course, if $\sqrt \Omega C_{abcd}$ is direction dependent but not
$\sqrt \Omega R_{ab}$, $\sqrt \Omega R_{abcd}$ will also fail to be
direction
dependent.)

Also, by the peeling property, $\Omega^{-1} C_{abcd}$ will exist on
$\cal I$.
This means that $\sqrt \Omega C_{abcd}$ is trivially zero on $\cal I$
and that the
condition that $\sqrt \Omega C_{abcd}$ be direction dependent should
be changed
there, cf.\ the situation for the electromagnetic field in section
\ref{remvidio}.

In a way, we can regard the metric as a potential for the Weyl
tensor. Thus
we prefer to look at the ``field'' rather than the ``potential''.

We will now recall the concepts of regularly direction dependent functions,
asymptotic flatness and direction dependent differentiable structures
of class $C^{>1}$, and also introduce the weaker structure of class
$C^{1^+}$. Thus definitions \ref{deff1}, \ref{C>1str} and \ref{svagdefasflat}
below are taken more or less from for instance
\cite{asht1} and \cite{wald}.

The starting point is a manifold which is smooth everywhere except at one
point $p$ where it is $C^1$ so that one at least has a tangent space there.
Thus it is possible to talk about directions at $p$. This enabels us to
talk about
direction dependent functions at p, i.e. functions $f$ for
which the limit along any $C^1$-curve $\gamma$ ending at $p$
exists and depends only on the tangent direction to $\gamma$ at $p$.
We write this as $\lim_{p}f=f(\eta)$ where $\eta$ is the tangent
vector to $\gamma$ at $p$.
We can then  define what it means for a function to be
regularly direction dependent:
\begin{definition} \label{deff1}
Let $M$ be a manifold which is $C^\infty$ everywhere except at a point $p$,
where it is $C^1$. Let $f$ be a function which is direction dependent at
$p$, and
let $(U,\Psi)$ be a chart
containing $p$, with coordinates $x^i$ so that $x^i(p)=0$. In terms of these
coordinates, let $F$ be constant on rays from the origin, so that
$\lim_{\Psi(p)}(f \circ \Psi ^{-1})=F(\eta)$. $f$ is then said to be regular
direction dependent (with respect to this chart) if, for all
$m \geq 0$, $1 \leq k \leq m$, $1 \leq i_k,j_k \leq 4$,
\beqn
\lim_{\Psi(p)}(x^{i_1}\frac{\p \phantom{x}}{\p x^{j_1}})...
(x^{i_m}\frac{\p \phantom{x}}{\p x^{j_m}})(f \circ \Psi ^{-1})=
[(x^{i_1}\frac{\p \phantom{x}}{\p x^{j_1}})...(x^{i_m}\frac{\p \phantom{x}}
{\p x^{j_m}})F](\eta) \label{regdirdep}
\eeqn
If $x^i(p)=\alpha^i$, $\alpha^i \neq 0$ for some $1 \leq i \leq 4$, $f$ is
said to be
regular direction dependent at $p$ if {\rm (\ref{regdirdep})} holds with
respect to
the translated coordinates $\widetilde x^i=x^i-\alpha^i$.
\end{definition}

If a function $f$ is regular direction dependent with respect to the chart
$(U,\Psi)$ above, the same need not be true for another chart (containing $p$)
in the $C^1$ atlas. For, let $(V,\Phi)$ be another chart with coordinates
$y^i$, $y^i(p)=0$ and suppose $f$ is the component, $f_k$, of a tensor field
$t_a=f_i(dx^i)_a=\tilde f_i(dy^i)_a$ say, so that $\tilde f_k=
f_l\frac{\partial x^l}{\partial y^k}$. Writing, for simplicity,
$x^i \pp {{x^j}} f_k$ instead of $(x^i \pp {x^j} (f_k \circ
\Psi^{-1}))\circ \Psi$,
we see that even if $x^i \pp {x^j} f_k$ is direction dependent for all $i$
and $j$, $y^i \pp {y^j} \tilde f_k=y^i \pp {y^j} (f_l\frac{\partial
x^l}{\partial y^k})=
(y^i \pp {y^j} f_l)\frac{\partial x^l}{\partial y^k}
+f_l y^i \pp {y^j} \frac{\partial x^l}{\partial y^k}
$
 need not be since we have no control
over $\lim_p \frac{\partial^2 x^l}{\partial y^k \partial y^j}$.
Thus the $C^1$ structure
is too ``large'' and we have to choose a smaller atlas, a $C^{>1}$ differential
structure. The following definition is similar to the one given in
\cite{asht1}.
\begin{definition}\label{C>1str}
Let $M$ be a manifold which is $C^\infty$ everywhere except at a point $p$,
where it is $C^1$. Let $(U,\Psi),(V,\Phi)$ be two charts containing $p$, with
coordinates $x^i$ and $y^i$ respectively. $(U,\Psi),(V,\Phi)$ are said to
be $C^{>1}$-related at $p$ if for all $i,j$ and $k$, both
\beqn
\frac {\partial^2 y^i }{\partial x^j \partial x^k}
\quad \mbox{and} \quad
\frac {\partial^2 x^i }{\partial y^j \partial y^k}
\eeqn
have regular direction dependent limits at $p$, in terms of the charts
$(U,\Psi)$ and $(V,\Phi)$ respectively.
A manifold $M$ is said to be $C^{>1}$ at a point $p$ if (it is otherwise smooth
and if ) all charts containing $p$ are $C^{>1}$-related.

A function f on $M \setminus \{p\}$ is said to be regularly direction
dependent at $p$
(where $M$ is assumed $C^{>1}$) if f is regular direction dependent in any
chart
containing~$p$, and we then write $f \in C^{>-1}(p)$.

A tensor field on $M \setminus \{p\}$ is said to be regularly
direction dependent at $p$ if its components in any chart are.
\end{definition}

We have not introduced any metric in the definition of a $C^{>1}$
manifold $M$. Since a (non-zero) vector field on a $C^k$ manifold is at
most only
$C^{k-1}$, the metric $g_{ab}$ on a manifold which is $C^{>1}$ at a point $p$
is at most $C^{>0}$ there. By this we mean that the first derivatives
of the components of the metric are regular direction dependent. In particular,
the metric is continuous at p, and since a $C^{>1}$ manifold is $C^1$, we have
a tangent space with metric at $p$. In the same way, any tensor field is
said to
be $C^{>0}$ at $p$ if it is continuous at $p$ and if the derivatives (with
respect to some chart in the $C^{>1}$ atlas) are regular direction dependent.
In order to give another motivation for the introduction of a weaker
differentiable structure, the $C^{1^+}$-structure, let us look again at
the properties of the $C^{>1}$-structure.\\
In order to ensure that the property
of beeing regularly direction dependent is independent of the choice
of charts, $x^i$ and $y^i$ say, we imposed the condition that
$
\frac {\partial^2 y^i }{\partial x^j \partial x^k}
$
and
$
\frac {\partial^2 x^i }{\partial y^j \partial y^k}
$
both be regularly direction dependent.
However, looking at the transformation rules for the components of a
tensor field, the primary requirement is that
$
\lim_{p} y^m\frac {\partial^2 y^i }{\partial x^j \partial x^k}=0
$
and
$
\lim_{p}x^m\frac {\partial^2 x^i }{\partial y^j \partial y^k}=0.
$

The $C^{1^+}$-condition will allow
$\lim_p \frac{\partial^2 x^l}{\partial y^k \partial y^j}$ to
diverge logarithmically at $p$, but keep the property
\beqn
\lim_p y^i\frac{\partial^2 x^l}{\partial y^k \partial y^j}=0.
\eeqn
\begin{definition} \label{C1pdeff}
Let $M$ be a manifold which is $C^\infty$ everywhere except at a point $p$,
where it is $C^1$. Let $(U,\Psi),(V,\Phi)$ be two charts containing $p$, with
coordinates $x^i$ and $y^i$ respectively.
Let {\rm (}via a translation if necessary {\rm )} $x^i(p)=0$, $y^i(p)=0$
and put
$\rho_x=\sqrt{(x^1)^2+...+(x^4)^2}$,
$\rho_y=\sqrt{(y^1)^2+...+(y^4)^2}$.
$(U,\Psi),(V,\Phi)$ are said to
be $C^{1^+}$-related at $p$ if for all $i$
\begin{eqnarray}
y^i	 & = & k^i(x) \rho_x^2 \log \rho_x + h^i (x)\rho_x^2+c^i_jx^j  \\
x^i	 & = & \tilde k^i(y) \rho_y^2 \log \rho_y + \tilde h^i
(y)\rho_y^2+\tilde c^i_jy^j
\end{eqnarray}

where each $k^i(\tilde k^i)$ is constant along rays, each $h^i(\tilde h^i)$
is regular direction
dependent {\rm (}in terms of the appropriate charts{\rm )} and the
$c_j^i(\tilde c_j^i)$
are constants.
A manifold $M$ is said to be $C^{1^+}$ at a point $p$ if {\rm (}it is
otherwise smooth
and if {\rm )} all charts containing $p$ are $C^{1^+}$-related.
\end{definition}

Regular direction dependent tensor field are defined as in definition
\ref{C>1str}.
Note that if we do not allow the logarithmic terms, we will get the
$C^{>1}$-structure.

Using a $C^{1^+}$-structure, (non-zero) tensor fields can, in general, be
at most $C^{0^+}$ at $i^0$.
By this we mean the following.
\begin{definition}\label{C0pdeff}
Let $M$ be a manifold which is $C^{1^+}$ at a point $p$.
Let ${T^{a...b}}_{c...d}$ be a tensor field on $M$.
${T^{a...b}}_{c...d}$ is said to be $C^{0^+}$ at $p$ if its component with
respect to any coordinates $x^i$, $x^i(p)=0$ {\rm (}in the $C^{1^+}$-atlas
{\rm)}
 can be
written
$$
{T^{i...j}}_{k...m}={k^{i...j}}_{k...m}\rho_x \log \rho_x +{h^{i...j}}_{k...m}
$$
where each ${k^{i...j}}_{k...m}$ is constant along rays and each
${h^{i...j}}_{k...m}$ is a $C^{>0}$-function.
\end{definition}
Even if the differentiable stucture at $p$ is only $C^{1^+}$, there
is an exceptional case where functions (and tensorfields) can have
higher regularity, namely if the function has a zero of sufficient
order at $p$. For instance, we say that a function $f$ is $C^{2^+}$ at $p$
if it, in some chart,  can be written as
\beqn
f(y)=c_{ij}y^i y^j+h(y)\rho_{y}^3+k(y)\rho_{y}^3 \log \rho_{y}
\eeqn
where the $c_{ij}$:s are constants and again $k$ is constant along
rays in the choosen chart and $h$ is a regularly direction dependent
function.\\
The definition of an asymptotically flat space-time will be the same as in
\cite{asht1} or \cite{wald} except that we use the $C^{1^+}$ structure.
Thus, citing \cite{wald}, we have

\begin{definition} \label{svagdefasflat}
A vacuum space-time $(M,g_{ab})$ is called asymptotically flat at null and
 spacelike infinity if there exists a space-time $(\widetilde M,\widetilde
g_{ab})$,
with $\widetilde g_{ab}$ $C^\infty$ everywhere except at a point $i^0$ where
$\widetilde M$ is $C^{1^+}$ and $ \widetilde g_{ab}$ is
$C^{0^+}$, and a conformal factor $\Omega$ satisfying the following conditions:
\begin{enumerate} \label{defasflat}

\item $\overline{J^+(i^0)} \cup \overline{J^-(i^0)}=\widetilde M \setminus M$
\item There exists an open neighbourhood V of $\dot M=\widetilde M \setminus M$
such that the space-time $(V,\widetilde g_{ab})$ is strongly causal.
\item $\Omega$ can be extended to a function on all of $\widetilde M$ which is
$C^{2^+}$ at $i^0$ and $C^\infty$ elsewhere.
\item (a) On ${\spp} = \dot{J} ^+(i^0) \setminus i^0$ and ${\sm}=\dot{J}
^-(i^0) \setminus i^0$ we have $\Omega = 0$ and $\widetilde \nabla_a \Omega
\neq 0$.
(b) We have $\Omega(i^0)=0$, $\widetilde \nabla _a \Omega(i^0)=0$, and
$\widetilde \nabla _a \widetilde \nabla _b \Omega(i^0)=-2\widetilde
g_{ab}(i^0)$.
\item (a) The map of null directions at $i^0$ into the space of integral curves
of $n^a = \widetilde g^{ab} \widetilde \nabla _a \Omega$ on $\spp$ and
$\sm$ is a
diffeomorphism.

(b) For a smooth function, $\omega$, on $\widetilde M \setminus i^0$
with $\omega > 0$ on $M \cup {\spp} \cup {\sm}$ which satisfies
$\widetilde \nabla_a(\omega^4 n^a)=0$ on ${\spp} \cup {\sm}$, the vector field
$\omega^{-1} n^a$ is complete on ${\spp} \cup {\sm}$.
\end{enumerate}
\end{definition}

For a detailed discussion of the meaning of these conditions, we refer
to \cite{wald} and \cite{asht1}.\\
In the next section we will look at the Bianchi identity near $i^0$.
Therefore we need to look at the curvature components connected to a
$C^{0^+}$ metric. So, let $\{x^i\}$ be the coordinates of some chart
with $x^i(i^0)=0$ and put $\rho^2=\sum_i (x^i)^2$. Then the components
of the metric are
\beqn
g_{ij}=\eta_{ij}+\rho\!\log\! \rho\; k_{ij}
\eeqn
where $\eta_{ij}$ are $C^{>0}$ functions
which have the Minkowski metric as limit at $i^0$ and where $k_{ij}$
are constant along rays (in terms of the coordinates) from $i^0$.
>From the definition of the Christoffel symbols it follows
that we can write
\beqn
{\Gamma^\rho}_{\mu \nu}=\log\!\rho \:{k^\rho}_{\mu
\nu}+{\gamma^\rho}_{\mu \nu},
\eeqn
where the ${k^\rho}_{\mu \nu}$'s are constant along rays and where
each
${\gamma^\rho}_{\mu \nu}$ is a regular direction dependent function,
and that the curvature
components can be
written as
\beqn
\rho{R_{\mu \nu \rho}}^\sigma=\log\!\rho {k_{\mu \nu
\rho}}^\sigma+{\gamma_{\mu \nu \rho}}^\sigma,
\eeqn
where again each ${k_{\mu \nu \rho}}^\sigma$ is constant along rays
and the
${\gamma_{\mu \nu \rho}}^\sigma$ are regular direction dependent
functions.
If this is true in one chart, it is true in all
(given the ``right'' differentiable structure).
Our aim is to prove that the logarithmic term is absent for the Weyl
tensor, i.e., that
\beqn
\rho{C_{\mu \nu \rho}}^\sigma={\gamma_{\mu \nu \rho}}^\sigma
\label{weylok}
\eeqn
for some regular direction dependent functions
${\gamma_{\mu \nu \rho}}^\sigma$.

It will be convenient to use the properties of the function $\rho$
above without
referring to a particular chart. Chosing another chart
$y^i,y^i(i^0)=0$, and
writing $\rho_y^2=(y^1)^2+...(y^4)^2$, we find, using the properties
in
Definition \ref{C1pdeff},
\beqn
\rho_y=k_1\rho (1+h\rho +k_2 \rho \log \rho), \label{rhofrihet}
\eeqn
where both $k_1$ and $k_2$ are constant along rays, and where h is a
regular
direction dependent function.
We will therefore often consider functions $r$ (on $M$) which can be
written
in the form (\ref{rhofrihet}), i.e., $r=k_1\rho (1+h\rho +k_2 \rho
\log \rho)$
in some (and therefore every) chart.
\section{The Bianchi identity near $i^0$}
\label{ubianchi}
In this section we will study the asymptotic Bianchi equation, i.e.
the Bianchi
equation near $i^o$.
We will assume that we have a space-time which is asymptotically flat
at null and
spacelike infinity in the sense of definition \ref{svagdefasflat}.
We saw in the previous section, that if we have a $C^{0^+}$ metric
at $i^0$, then the
curvature tensor ${R_{abc}}^d$ behaves, near $i^0$, as
\beqn
\rho{R_{abc}}^d=\log\!\rho \,{k_{abc}}^d+{h_{abc}}^d,
\eeqn
where ${k_{abc}}^d$ and
${h_{abc}}^d$
are two regular direction dependent tensor fields with ${k_{abc}}^d$
constant along rays.
The same is of course true for the Weyl tensor ${C_{abc}}^d$. We will
show that,
for the Weyl tensor, there is no logarithmic term, i.e., we have
$r{C_{abc}}^d={h_{abc}}^d$ near $i^0$. In fact we will show that
\beqn
r^3\frac{{C_{abc}}^d}{\Omega}={h_{abc}}^d, \label{attvisa}
\eeqn
where $r$ is any function satisfying (\ref{rhofrihet}) and where
$\Omega$ is the
conformal factor.
Instead of using the Weyl tensor ${C_{abc}}^d$, we will use the Weyl
spinor,
and in order to show (\ref{attvisa}) we will need a direction
dependent null tetrad and spin
frame near $i^o$. As shown in appendix A, one can choose a direction
dependent
null tetrad $N^a,L^a,M^a$ and an associated spin frame which has
essentially the following properties: If we denote standard polar
coordinates in Minkowski space-time with $t,r,\theta,\varphi$ we can
think of our tetrad, in the limit, as
\beqn
\! N^a\!&\!\!\!\!=&\!\!\!\!\!\frac{1}{\sqrt 2}(\pp t + \pp r)^a,
\quad
\! L^a\!=\!\frac{1}{\sqrt 2}(\pp t - \pp r)^a, \nonumber \\
\! M^a\!&\!\!\!\!=&\!\!\!\!\!\frac{1}{\sqrt 2 r}(\pp \theta -
\frac{i}{\sin \theta}\pp \varphi)^a. \label{astetrad}
\eeqn
One can also choose a radial parameter r, essentially (i.e. in the
limit)
the ``polar'' r above such that
\beqn
\lim_{i^o}\nabla_a r=-N_a+L_a.	\label{rvillkor}
\eeqn
We want to adopt the GHP formalism, \cite{ghp}, \cite{pen1}, so we
now let
$(n^a,l^a,m^a,\bar{m}^a)$ be weighted vector fields (with weights
\{-1,-1\},\{1,1\},\{1,-1\},\{-1,1\} respectively), with $n^a,l^a$
pointing along the directions of $N^a,L^a$, and let $(o^A,\iota^A)$
be the spinor dyad
corresponding to \newline
$(n^a,l^a,m^a,\bar{m}^a)$.
We also define the weighted scalar $e$, which we require to be
$C^{>0}$, of weight
\{1,1\} by
\beqn
e=l^aN_a.
\eeqn
Thus we have, at $i^0$,
\beqn
\nabla_a r=-N_a+L_a=-en_a+e^{-1}l_a.\label{nablar}
\eeqn
and therefore also
\beqn
\lim_{i^0}m^a\nabla_ar=\lim_{i^0}\overline  m^a\nabla_ar=0,\;
\nonumber \\
\lim_{i^0}l^a\nabla_ar=-e,\quad \lim_{i^0}n^a\nabla_ar=e^{-1}.
\eeqn
Let us denote the space of spacelike and null directions at $i^0$ by
$K$, so
that $k \in K$ is any such direction. We then define derivative
operators
on $K$ as follows. Let $\eta^o \in C^{>-1}(i^0)$, i.e., $\eta^o$ is
any regularly
direction dependent quantity, and put
\beqn
\eta(k)=\lim_{x \to i^0} \eta^o(x).
\eeqn
We then define the derivative operator $\partial_a$ on $K$ by
\beqn
\partial_a \eta(k) = \lim_{x \to i^0} r \nabla_a \eta^o(x)
\label{derdeff}
\eeqn
Note that although $\partial_a$ depends on the choice of function $r$,
$\partial_a$ is independent of the components of the Christoffel
symbol,
${\Gamma^i}_{jk}$ since
$\lim_{i^0} r{\Gamma^i}_{jk}=0$.
Thus we can use a coordinate derivative on the RHS
of (\ref{derdeff}) and get the same results as in flat space-time.

In the GHP formalism, we now denote the usual derivative operators by
$\Ph^o$, $\Ph^{\prime o}$, $\D^o$ and $\D^{\prime o}$, so that e.g.,
on a
zero-weighted scalar $\eta$, $\Ph^o \eta=l^a\nabla_a\eta$ etc.
We then define, for any weighted $C^{>-1}(i^0)$ scalar $\eta^o$ the
following
derivative operators on $K$.
\beqn
\Ph \eta(k)=\lim_{i^0}r\Ph^o\eta^o, \quad\Ph'
\eta(k)=\lim_{i^0}r\Ph^{\prime o}\eta^o
\nonumber \\
\D \eta(k)=\lim_{i^0}r\D^o\eta^o, \quad\D'
\eta(k)=\lim_{i^0}r\D^{\prime o}\eta^o.
\eeqn
Using these derivative operators we find, in the limit at $i^o$,
(again we refer
to appendix A) i.e. on K that
\begin{equation}
\Ph o_A=\Php o_A = \Ph \iota_A=\Php \iota_A = \D o_A =\Dp \iota_A =
0\label{spinder1}
\end{equation}
\begin{equation}
	      \D \iota_A =  \ei o_A \label{spinder2} \quad
\end{equation}
	      \begin{equation}
		      \Dp o_A =-e \iota_A
	      	\label{}
	      \end{equation}
	      \begin{equation}
		      \Ph e= \Php e=\D e=\Dp e=0.\label{spinder3}
	      \end{equation}
and that
\begin{eqnarray}
		\lim_{i^0}r\rho=e, \quad \lim_{i^0}r\rho'=-e^{-1} \\
	\lim_{i^0}r\kappa=\lim_{i^0}r\kappa'=\lim_{i^0}r\sigma=
	\lim_{i^0}r\sigma'=\nonumber \\
	\lim_{i^0}r\tau=\lim_{i^0}r\tau'=
	\lim_{i^0}r\epsilon=\lim_{i^0}r \epsilon'=0.
\end{eqnarray}
The commutator relations become simply
\beqn
\Ph\Php-\Php\Ph&\!\!\!\!=&\!\!\!\!-e\Php-\ei\Ph,\\
\Ph\D-\D\Ph&\!\!\!\!=&\!\!\!\!0, \\
\D\Dp-\Dp\D&\!\!\!\!=&\!\!\!\!(p-q),
\eeqn
plus the primed and conjugated versions.

Let us finally derive another expression for the operators $\Ph$ and
$\Php$
on $K$. Again referring to the polar coordinates $t,r,\theta,\varphi$
in the tangent space at $i^o$, we
may label the directions in $K$ with $\alpha,\theta,\varphi$ where
\beqn
\alpha=\frac{t}{r}.
\eeqn

Thus $\alpha=1\;(\alpha=-1)$ for future (passed) directed null
directions, and on
K one finds that
\beqn
\Ph=e(1+\alpha)\pa, \label{phalfa}\\
\Php=\ei(1-\alpha)\pa \label{phpalfa}.
\eeqn

We now return to the equation of interest, i.e. the Bianchi equation.
We put
\beqn
 \varphi_{ABCD}=\Omega^{-1}\Psi_{ABCD}.
\eeqn
By virtue of the peeling property, $\varphi_{ABCD}$ exists on $\cal
I$.
>From the discussion in the beginning of this section, we know that,
near $i^0$,
\beqn
r\Psi_{ABCD}=\log\!r\,k_{ABCD}+h_{ABCD},
\eeqn
where r is of type (\ref{rhofrihet}), and $k_{ABCD}$, $h_{ABCD}$ are
regular
 direction dependent (symmetric) spinor fields, where the components
of $k_{ABCD}$ can be
taken to be constant along rays (in the chosen spin frame).
>From the definition of a $C^{2^+}$-function it follows that $\Omega$
has the property that for any chart (within the $C^{1^+}$-structure)
$\Omega$ takes the form $\Omega=r^2\,k$, where $r$ satisfies
(\ref{rvillkor}) and $k$ is a function which is constant
along rays and which vanishes on $\cal I$.
Thus,
\beqn
r^3 \frac{1}{\Omega}\Psi_{ABCD}=r^3
\varphi_{ABCD}=\log\!r\,\mu_{ABCD}+\gamma_{ABCD}.\label{fimygamma}
\eeqn
Again, $\mu_{ABCD}$ and $\gamma_{ABCD}$ are regular direction
dependent
symmetric spinor fields where the components can be taken as for
$k_{ABCD}$
and $h_{ABCD}$.

We will now show that $\mu_{ABCD}=0$.

Let us first recall that, with $\widehat \Psi_{ABCD}$ being the Weyl
spinor in
physical space-time $\widehat M$, the Bianchi identity reads
\beqn
\widehat \nabla^{AA'}\widehat \Psi_{ABCD}=0.
\eeqn
Here $\widehat \nabla_a$ is the derivative operator associated with
the physical
metric $\widehat g_{ab}$. Under a conformal rescaling $\widehat
g_{ab} \to g_{ab}=
\Omega^2\widehat g_{ab}$, the Weyl spinor is conformally invariant,
\beqn
\Psi_{ABCD}=\widehat \Psi_{ABCD}.
\eeqn
For a proof of this, see \cite{pen2}.

In the rescaled space-time $M$, with derivative operator $\nabla_a$,
the Bianchi
identity is
\beqn
\nabla^{AA'}(\frac{1}{\Omega}\Psi_{ABCD})=0.
\eeqn
Therefore with the notation adopted, we have
\beqn
\nabla^{AA'}\varphi_{ABCD}=0.\label{fibianchi}
\eeqn
>From (\ref{fimygamma}) we thus have
\beqn
\mu_{ABCD}=\frac{r^3}{\log r}\varphi_{ABCD}-\frac{1}{\log
r}\gamma_{ABCD}.
\label{myfigamma}
\eeqn
We differentiate this equation, multiply with $r$ and get
\beqn
r\nabla^{AA'}\mu_{ABCD}&\!\!\!\!=&\!\!\!\!
\frac{3r^3}{\log r}(1-\frac{1}{3\log r})\varphi_{ABCD}
\nabla^{AA'}r\nonumber \\
&&\!\!\!\!+\frac{r^4}{\log r}\nabla^{AA'}\varphi_{ABCD}-
r\nabla^{AA'}(\frac{\gamma_{ABCD}}{\log r}).
\eeqn
Taking the limit, the last term on the RHS will vanish, and by using
(\ref{fibianchi}) and (\ref{myfigamma}) we get the following equation
on $K$
\beqn
\p^{AA'}\mu_{ABCD}=3\nabla^{AA'}\!r\;\mu_{ABCD}, \label{mubianchi}
\eeqn
where, from (\ref{nablar}),
\beqn
\nabla^{AA'}\!r=-e\iota^A\iota^{A'}+\ei o^Ao^{A'}\label{spinnablar}
\eeqn
at $i^0$.
In order to solve (\ref{mubianchi}) (and (\ref{fibianchi})) we make
the usual
decomposition of $\mu_{ABCD}$ and $\gamma_{ABCD}$, so that
\begin{eqnarray}
\mu_{0}=\mu_{ABCD}o^Ao^Bo^Co^D,\;\;&\gamma_{0}
&=\gamma_{ABCD}o^Ao^Bo^Co^D \\
\mu_{1}=\mu_{ABCD}\,\iota^Ao^Bo^Co^D,\;\;&\gamma_{1}
&=\gamma_{ABCD}\,\iota^Ao^Bo^Co^D \\
\mu_{2}=\mu_{ABCD}\,\iota^A\,\iota^Bo^Co^D,\;\;&\gamma_{2}
&=\gamma_{ABCD}\,\iota^A\,\iota^Bo^Co^D \\
\mu_{3}=\mu_{ABCD}\,\iota^A\iota^B\,\iota^Co^D,\;\;&\gamma_{3}
&=\gamma_{ABCD}\,\iota^A\iota^B\,\iota^Co^D \\
\mu_{4}=\mu_{ABCD}\,\iota^A\,\iota^B\iota^C\,\iota^D,\;\;&\gamma_{4}
&=\gamma_{ABCD}\,\iota^A\,\iota^B\iota^C\,\iota^D \\
\end{eqnarray}
Let us now look at the LHS of equation (\ref{mubianchi}). We have
\begin{eqnarray}
\label{derivation}
\p^{AA'}\mu_{ABCD}&=&\lim_{i^0}\epsilon^{AE}r
\nabla_{\!E}^{\;A'}\mu_{ABCD}\nonumber
\\
&=&\lim_{i^0}(o^A\iota^E-\iota^Ao^E)r
\nabla_{\!E}^{\;A'}\mu_{ABCD}.
\end{eqnarray}
So, transvecting with $o_{A'}$ and $\iota_{A'}$ respectively, and
using
(\ref{spinnablar}), we obtain
\begin{eqnarray}
(o^A\Dp-\iota^A\Ph)\mu_{ABCD}&=& 3e\iota^A\mu_{ABCD} \\
(o^A\Php-\iota^A\D)\mu_{ABCD}&=& 3\ei o^A\mu_{ABCD}.
\end{eqnarray}
We now contract these equations with $o^Bo^Co^D$, $o^Bo^C\iota^D$,
$o^B\iota^C\iota^D$ and $\iota^B\iota^C\iota^D$. Together with
Leibniz' rule and
(\ref{spinder1})-(\ref{spinder3}), we get
\begin{eqnarray}
\Ph \mu_{1}-\Dp \mu_{0}-e \mu_{1}&=&0, \label{phmu1} \\
\Php \mu_{0}-\D \mu_{1}-2\ei \mu_{0}&=&0,\\
\Ph \mu_{2}-\Dp \mu_{1}&=&0, \label{phmu2}\\
\Php \mu_{1}-\D \mu_{2}-\ei \mu_{1}&=&0,\\
\Ph \mu_{3}-\Dp \mu_{2}+e \mu_{3}&=&0,\\
\Php \mu_{2}-\D \mu_{3}&=&0, \label{phpmu2}\\
\Ph \mu_{4}-\Dp \mu_{3}+2e \mu_{4}&=&0,\\
\Php \mu_{3}-\D \mu_{4}+\ei \mu_{3}&=&0.\phantom{.} \label{phpmu3}
\end{eqnarray}
Note that $\Php$ gives zero when acting on directions on $\cal I^+$,
so that
equation (\ref{phpmu2}) gives $\D \mu_{3}(k)=0$, $k \in \cal I^+$.
Since $\mu_{3}$
has spin weight -1, this implies $\mu_{3}(k)=0$, $k \in \cal I^+$,
\cite{pen1}.
Applying this result to (\ref{phpmu3}), we find, since $\mu_{4}$ has
negative
spin weight, that also $\mu_{4}(k)=0$, $k \in \cal I^+$.
The same conclusion holds for $\mu_{0}$ and $\mu_{1}$ on $\cal I^-$.
This means that $\mu_{2}(k)$, $k \in \cal I^+$ is unaffected by the
change of
spin basis $(o^A,\iota^A) \to (o^A+\lambda \iota^A,\iota^A)$ (and that
$\mu_{2}(k)$, $k \in \cal I^-$ is invariant under $(o^A,\iota^A) \to
(o^A,\iota^A+\lambda o^A)$). It is also clear that $\mu_{0}$,
$\mu_{1}$,
$\mu_{3}$ and $\mu_{4}$ are determined by $\mu_{2}$.

Note that the above equations come naturally in pairs. Using this and
the
commutator relations we derive the following equations.
\begin{eqnarray}
(\Ph \Php -2 \ei \Ph -e\Php +2-\D\Dp)\mu_{0}&=&0, \label{mu0leg}\\
(\Ph \Php - \ei \Ph -\D\Dp)\mu_{1}&=&0, \\
(\Php \Ph - \ei \Ph -\D\Dp)\mu_{2}&=&0, \label{mu2leg}\\
(\Php \Ph + e \Php -\Dp\D)\mu_{3}&=&0, \\
(\Php \Ph +2 e \Php +\ei\Ph +2-\Dp\D)\mu_{4}&=&0. \label{mu4leg}
\end{eqnarray}
We now use (\ref{phalfa}) and (\ref{phpalfa}) on (\ref{mu2leg}). We
get
\beqn
((1-\al^2) \pat - 2 \al \pa - \D \Dp ) \mu_{2}=0 \label{mu2legalfa}.
\eeqn
Writing the directions $k$ as $k=(\al,\theta,\varphi)$, where
$(\theta,\varphi)$
are standard spherical coordinates (and $-1 \leq \al \leq 1$), we
expand, for
each $\al$, $\mu_{2}$ in spherical harmonics. Thus
\beqn
\mu_{2}(\al,\theta,\varphi)=\sum_{n=0}^{\infty}\sum_{m=-n}^{n}c_n^m
(\al)Y_n^m (\theta,\varphi),
\eeqn
where  each $c_n^m(\al)$ becomes smooth since $\mu_{2}$ is assumed
smooth.
Using that $\D \Dp Y_n^m=$ $ -n(n+1)Y_n^m$ and that the series may be
 differentiated termwise, we get the following equations for the
$c_n^m$'s.
\beqn
\left((1-\al^2) \pat - 2 \al \pa +n(n+1) \right) c_n^m=0,
\eeqn
i.e., the Legendre differential equation.\\
For each $n \in \bf N$, we have the Legendre polynomial  $p_n$ as
solution.
The Legendre functions of the second kind are not allowed since they
diverge
at $\al =1$ and $\al = -1$. Note that this is an essential restriction
compared to \cite{asht1}.

Let us write
$\mu_{2}(1,\theta,\varphi)=\sum_{n=0}^{\infty}\sum_{m=-n}^{n}d_n^mY_n^m(\theta,\
varphi)$.
Since $p_n(1)=1$ for all $n$, we can write
$\mu_{2}$ as
\beqn
\mu_{2}(\al,\theta,\varphi)=\sum_{n=0}^{\infty}\sum_{m=-n}^{n}d_n^mp_n(\al)Y_n^m
(\theta,\varphi).
\label{mu2expan}
\eeqn
We now return to equation (\ref{fimygamma}). We differentiate this
relation,
multiply with $r$, use (\ref{fibianchi}) and (\ref{fimygamma}) again
and
find
\beqn
\log
r\{3\nabla^{AA'}\!r\;\mu_{ABCD}-r\nabla^{AA'}\mu_{ABCD}\}=\nonumber \\
r\nabla^{AA'}\gamma_{ABCD}-3\nabla^{AA'}\!r\;\gamma_{ABCD}+\nabla^{AA'}\!r\;\mu_
{ABCD}
\label{nastord}
\eeqn
By (\ref{mubianchi}), the bracket on the LHS of (\ref{nastord}) is,
in the limit,
 zero. Furthermore, by choosing $\mu_{ABCD}$ such that the components
in some
dyad are constant along rays, we conclude that the LHS behaves like
$\log r\,\rho \log \rho$, i.e., goes to zero as $r \to 0$.
Thus, taking the limit $r \to 0$, we get the equation
\beqn
3\nabla^{AA'}\!r\;\gamma_{ABCD}-\p^{AA'}\gamma_{ABCD}=
\nabla^{AA'}\!r\;\mu_{ABCD}.
\eeqn
A calculation, analogous to the above one yields
\begin{eqnarray}
\Ph \gamma_{1}-\Dp \gamma_{0}-e
\gamma_{1}\!&=&\!-e\mu_1,\label{phgamma1}\\
\Php \gamma_{0}-\D \gamma_{1}-2\ei \gamma_{0}\!&=&\!\ei\mu_0,
\label{phpgamma0}\\
\Ph \gamma_{2}-\Dp \gamma_{1}\!&=&\!-e\mu_2, \label{phgamma2}\\
\Php \gamma_{1}-\D \gamma_{2}-\ei \gamma_{1}\!&=&\!\ei\mu_1,
\label{phpgamma1}\\
\Ph \gamma_{3}-\Dp \gamma_{2}+e \gamma_{3}\!&=&\!-e\mu_3,\\
\Php \gamma_{2}-\D \gamma_{3}\!&=&\!\ei\mu_2, \label{phpgamma2}\\
\Ph \gamma_{4}-\Dp \gamma_{3}+2e
\gamma_{4}\!&=&\!-e\mu_4,\label{phgamma4}\\
\Php \gamma_{3}-\D \gamma_{4}+\ei \gamma_{3}\!&=&\!\ei\mu_3.
\label{phpgamma3}
\end{eqnarray}
Similarly we apply $\Php$ to (\ref{phgamma2}), $\Dp$ to
(\ref{phpgamma1}), use (\ref{phmu2}) and (\ref{phgamma2}) again. We
obtain
\beqn
(\Php \Ph -\ei \Ph - \D\Dp)\gamma_2=\mu_2-e \Php \mu_2 +\ei \Ph \mu_2.
\eeqn
We expand $\gamma_2$ in spherical harmonics, so that
$\gamma_{2}(\al,\theta,\varphi)$ takes the form \\
$\sum_{n=0}^{\infty}\sum_{m=-n}^{n}a_n^m(\al)Y_n^m(\theta,\varphi)$.
We also use the expansion (\ref{mu2expan}) for $\mu_2$.
Since $1-e \Php+\ei \Ph=1+2\alpha \pp \alpha$, $a_n^m$ must satisfy
\beqn
\{(1-\al^2)\dpat-2\al \dpa +n(n+1)\}a_n^m(\al)= \nonumber \\
d_n^m\{p_n(\al)+2\al\dpa p_n(\al)
\} \label{anmekv}.
\eeqn
Let us look for solutions of the type $a_n^m=v_n^mp_n$.
The resulting equation can be written
\beqn
\dpa\{(1-\al^2)p_n^2 \dpa v_n^m\}=d_n^m \dpa(\al p_n^2)
\eeqn
which implies that
\beqn
a_n^m=v_n^mp_n=-\frac{d_n^mp_n}{2}\log (1-\al^2)+C q_n+D p_n
\label{gamkoeff}
\eeqn
for some constants $C$ and $D$. Since $\gamma_{ABCD}$ is assumed
smooth, this
implies that $a_n^m$ must also be smooth.
We note that $\log(1-\al^2)$ diverges logarithmically when
$\al \to 1$ and $\al \to -1$.
$q_n$ also behaves in this way, but since $p_n$ and $q_n$ has
the opposite parity, $a_n^m$ can be defined for $\al=1$ and $\al=-1$
only
if $d_n^m$ and $C$ are zero. This means that
$\lim_{i^0}\mu_{ABCD}=0$, and,
by the properties of $\mu_{ABCD}$, that $\lim_{i^0}\log
r\mu_{ABCD}=0$.

We can now state the following theorem.
\newtheorem{sats}{Theorem} \label{antipod}
\begin{sats}
Suppose $(\widehat M,\widehat g_{ab})$ is asymptotically flat at null
and
spacelike infinity in the sense of Definition \ref{svagdefasflat},
with the rescaled metric
$C^{0^+}$ at $i^0$.
If $r$ is any $C^{>0}$ function with $r(i^0)=0$,
${{\bf C}_{abc}}^d(\eta)=\lim_{i^0} \frac{r^3}{\Omega}{{{\widetilde
C}_{abc}}}^{\phantom {ddd}d}$
exists and has the symmetry property ${{\bf C}_{abc}}^d(\eta)={{\bf
C}_{abc}}^d(-\eta)$ where $\eta$ is any  spacelike or null vector in
$T_{i^0}M$.
\end{sats}
Using the direction dependent tetrad $L^a,N^a,M^a,\overline M^a$ from
appendix A so that $L^a(\eta)=N^a(-\eta), M^a(\eta)=
\overline M^a(-\eta)$, and also using our decomposition of the Weyl
tensor,
we have to
show that
\beqn
\gamma_2(\eta)=\gamma_2(-\eta),\,\gamma_3(\eta)=\gamma_1(-\eta),\,\gamma_4(\eta)
=
\gamma_0(-\eta).
\eeqn
To show that $ \gamma_2(\eta)=\gamma_2(-\eta) $, we note that from
(\ref{gamkoeff}) with $d^m_n=C=0$,
we have
\beqn
\gamma_{2}(\al,\theta,\varphi)=\sum_{n=0}^{\infty}\sum_{m=-n}^{n}b_n^mp_n(\al)Y_
n^m(\theta,\varphi).
\label{gamma2expan}
\eeqn
Therefore, the symmetry property follows from
\beqn
 p_n(\alpha)=(-1)^n p_n(-\alpha) \label{legendreantipod}\\
 Y_n^m(\pi-\theta,\varphi \pm \pi)=
(-1)^nY_n^m(\theta,\varphi),\label{yklantipod}
\eeqn
 That $\gamma_3(\eta)=\gamma_1(-\eta)$ then follows from the symmetry
of the equations (\ref{phgamma2}) and
(\ref{phpgamma2}) (with $\mu_{ABCD}=0$). Similarly (\ref{phgamma1})
and
(\ref{phpgamma3}) give $\gamma_4(\eta)=\gamma_0(-\eta)$.

Note that using the
alternative definitions ${{\bf C}_{abc}}^d(\eta)=\lim_{i^0}
\Omega^{1/2}{{{\widetilde C}_{abc}}}^{\phantom {ddd}d}$ or ${{\bf
C}_{abc}}^d(\eta)=\lim_{i^0} r{{{\widetilde C}_{abc}}}^{\phantom
{ddd}d}$ , we still have that the limits exist and that ${{\bf
C}_{abc}}^d(\eta)={{\bf C}_{abc}}^d(-\eta)$,
$\eta \in T_{i^0}M$.

To see this, we note that the factor $\frac{\sqrt \Omega}{r}$ has the
symmetry
property $\lim_{i^0}\frac{\sqrt \Omega}{r}(\eta)=
\lim_{i^0}\frac{\sqrt \Omega}{r}(-\eta)$ which is obvious for null
directions
and which follows in spacelike directions if we let $\eta$ be
indicated by the
corresponding timelike unit vector $\eta^a$ at $T_{i^0}$ and use
$\lim_{i^0}\frac{\sqrt \Omega}{r}(\eta)=
\lim_{i^0}\frac{\eta^a \nabla_a \sqrt \Omega}{\eta^a \nabla_a
r}(\eta)=
\lim_{i^0}\frac{\eta^a \eta_a}{\eta^a (L_a-N_a)}$ which is invariant
under
$\eta^a \to - \eta^a$. Here again we have used the same tetrad as
above, so
that $(L_a-N_a)(\eta)=-(L_a-N_a)(-\eta)$.
\section{The electromagnetic field near $i^0$} \label{remvidio}
In this section we will discuss the difference in the
differentiability conditions
at $i^o$ by comparing with the corresponding conditions on the
electro-magnetic
field. In \cite{asht1}, with $\Omega$ being the comformal factor and
$F_{ab}$
the electromagnetic field, the imposed condition is that $\Omega
F_{ab}$ is
regular direction dependent at $i^o$. As we will argue, this
condition is too
weak, i.e. it is neccesary to include suitable requirements on $\spp$
and $\sm$.
In this way, we again motivate the general importance of
including the (limits at $i^0$ of ) fields on $\spp$ and $\sm$.

Consider the three pairs
\beqn
E&\!\!\!\!=&\!\!\!\!-\frac{dr}{r^2}, \phantom{AAAAAAAAAAAAA} B=0,
\label{punktkallaimink}\\
E&\!\!\!\!=&\!\!\!\!-\frac{t\cos
\theta}{r^3}dr-\frac{1}{2}\frac{t\sin \theta}{r^3}r\,d\theta,\quad
\displaystyle B=-\frac{1}{2}\frac{\sin \theta}{r^2} r \sin \theta
d\phi, \label{dipolimink}
\eeqn
\vspace{-3mm}
\beqn
E&\!\!\!\!=&\!\!\!\!-(\frac{t}{2r}\log(\frac {r+t}{r-t})-1)
\frac{2\cos \theta}{r^2}dr-(\frac{t}{2r}\log(\frac
{r+t}{r-t})-\frac{t^2}{t^2-r^2}) \frac{\sin \theta}{r^2}r\,d \theta,
\nonumber \\
B&\!\!\!\!=&\!\!\!\!-(\frac{1}{2}\log(\frac
{r+t}{r-t})-\frac{rt}{t^2-r^2}) \frac{sin \theta}{r^2}r\sin\theta \,d
\varphi.\hspace{20pt} \label{laddskalimink}
\eeqn
These are solutions to Maxwell's equation in flat space-time,
obtained in the
following way. By making a standard conformal completion of the
Minkowski
space-time and
taking the limit at $i^0$, in spacelike directions, to Maxwells
equation using
$\Omega F_{ab}$, one get an equation on $\widetilde K$, the
unit timelike hyperboloid in the tangent space at $i^0$. One get a
family of solutions, essentially parametrized by the Legendre
polynomials of the first and second
kind. By taking the first few solutions on $\widetilde K$, we write
down the corresponding solutions in our original spacetime.

The solution (\ref{punktkallaimink}) is just the field from a point
charge (placed
at $r=0$). The field (\ref{dipolimink}) can be thought of as a dipol
where the
dipol moment increases (linearly) with time. This may not seem very
physical,
and one can exclude, see \cite{avhandling}, such solutions by
imposing suitable conditions at timelike
infinity, $i^+$. The solution (\ref{laddskalimink}) is singular at
the light cone
of the origin, $t=0, r=0$. It may be thought of as a charged sphere,
expanding
with the speed of light. We believe that this is also not a very
physical
solution, and by changing the conditions on the electromagnetic field
near $i^0$,
solutions of this type will be excluded.

We can equally well consider the fields (\ref{punktkallaimink})-
(\ref{laddskalimink}) as fields in the tangent space at $i^0$, or in
this
case, as fields in the rescaled space-time by interpreting
$t,r,\theta,\varphi$
as polar coordinates with respect to $i^o$. In the standard
completion of
the Minkowski space-time, the rescaled space-time is again flat and
the conformal
factor $\Omega$ is related to the polar ccordinates above via
$\Omega=r^2-t^2$.

Thus in any spacelike direction, the condition that $\lim_{i^0}\Omega
F_{ab}$
exists as a regular direction dependent tensor field at $i^0$ is
equivalent
to the same condition on $\lim_{i^0}r^2 F_{ab}$, since
$\Omega=r^2(1-\frac{t^2}{r^2})$ and the limit of
$(1-\frac{t^2}{r^2})$ along any
spacelike direction is non-zero. On $\spp$ and $\sm$, the fields
(\ref{punktkallaimink}) and (\ref{dipolimink}) diverge like $1/r^2$
so that
$r^2 F_{ab}$ has a non-zero limit along null directions.
$\Omega F_{ab}$ is trivially zero on $\cal I$ for these fields and
imposes no restriction. Moreover, by continuity, $\Omega F_{ab}$
exists
and is zero on $\cal I$ also for the field given in
(\ref{laddskalimink}), but, for this
field, $R^2 F_{ab}$ is not even defined there.

We conclude that, even if the condition imposed on $\Omega F_{ab}$ is
very
natural in that it is completely coordinate independent, it might not
be
restrictive enough. A limiting condition on $R^2 F_{ab}$ seems more
appropriate,
but is, as it stands, formulated in coordinates. Below we will
suggest a coordinate independent version.

Let us also note that solutions corresponding to pairs $p_k,\;Y_k^l$
for higher
$k$, will all correspond to multipole moments with strength that
increases in
time, while solutions corresponding to $q_k,\; Y_k^l$ will all be
singular on $\cal I$.

Also, in exactly the same way, we can start with a magnetic field on
$K$ and
derive corresponding fields in $M$.

In the case of a general asymptotically flat space-time,
we
consider electromagnetic fileds $F_{ab}$ with the following
properties. For any
$C^{>0}$ function $r$ with $r(i^0)=0$, the field $r^2F_{ab}$  should
be regularly
direction dependent. We may analyse the situation by deriving the
equations corresponding to
(\ref{phmu1})-(\ref{phpmu3}) and (\ref{mu0leg})-(\ref{mu4leg}). We
use the tetrad from
appendix A, and put $\phi_{AB}=r^2\varphi_{AB}$, where
$\varphi_{AB}$
is defined by
$F_{ab}=\varphi_{AB}\overline{\epsilon}_{A'B'}+\epsilon_{AB}
\overline{\varphi}_{A'B'}$. Using the source free Maxwell equation,
$\nabla^{AA'}\varphi_{AB}=0$ and (\ref{derdeff}), we obtain, after a
derivation
analogous to the derivation on page \pageref{derivation}, the
following
equations at~$i^0$.
\begin{eqnarray}
\Ph \phi_1-\Dp\phi_0&=&0, \\
\Php \phi_0-\D \phi_1-\ei \phi_0&=&0, \\
\Ph \phi_2 -\Dp \phi_1 +e \phi_2&=&0, \\
\Php \phi_1 -\D \phi_2&=&0.
\end{eqnarray}
and the decoupled equations
\begin{eqnarray}
(\Ph \Php -\ei  \D \Dp)\phi_0&=&0,\\
(\Php \Ph - \ei \Ph - \Dp \D)\phi_1&=&0, \label{phi1leg}\\
(\Php \Ph + e \Php - \Dp \D)\phi_2&=&0.\!
\end{eqnarray}
We see that equation (\ref{phi1leg}) is the same as (\ref{mu2leg}).
Therefore
$\phi_1$ takes the form
\beqn
\phi_{1}(\al,\theta,\varphi)=\sum_{n=0}^{\infty}\sum_{m=-n}^{n}b_n^mp_n(\al)Y_n^
m(\theta,\varphi)
\eeqn
for some constants $b_n^m$.
Contributions from the Legendre functions of the second kind are
excluded by the
conditions on $r^2F_{ab}$ on $\spp$ and $\sm$. We also see that
$\phi_1$ has
the same antipodal property as $\gamma_2$ in the gravitational case.

Let us remark that we can also consider fields which behave like the
field
$r^3\varphi_{ABCD}$ in (\ref{fimygamma}), i.e., the situation when
\beqn
r^2\varphi_{AB}=\log \!r \, \mu_{AB}+\gamma_{AB}, \label{emlog}
\eeqn
where $\mu_{AB}$ and $\gamma_{AB}$ have the same properties as the
corresponding
fields in (\ref{fimygamma}).

By the results of the previous section, we expect that $\mu_{AB}$ can
be
taken smooth and non-zero at the expense of $\gamma_{AB}$ blowing up
at
$\spp$ and $\sm$. As an example of such a field, with standard
coordinates
$(t,r,\theta,\varphi)$ in Minkowski space-time, we take
\begin{eqnarray}
E&=&\frac{\log r}{r^2}(\frac{t}{r}\cos \!\theta \;\hat{\bf
r}+\frac{1}{2}
\frac{t}{r}\sin \!\theta \;\hat { {\!\theta}})+ \nonumber \\
&&\frac{1}{2}\frac{t}{r^3}(\log(1-\frac{t^2}{r^2})\cos\! \theta
\;\hat{\bf r}+
(-1+\frac{1}{2}\log(1-\frac{t^2}{r^2})-\frac{t^2/r^2}{1-t^2/r^2})\sin
\!\theta
\;\hat { {\!\theta}}, \nonumber \\
B&=&\frac{1}{2}\frac{\log r}{r^2}\sin \!\theta \;\hat {
{\!\varphi}}+
 \frac{1}{2}\frac{1}{r^2}(\frac{1}{2}\log(1-\frac{t^2}{r^2}))-
\frac{t^2/r^2}{1-t^2/r^2}))\sin \!\theta \;\hat { {\!\varphi}}.
\label{logexempelem}
\end{eqnarray}
We see that the logarithmic part of both $E$ and $B$ have, suitable
rescaled,
regular direction dependent limits at the origin, but that the next
order
behaviour, i.e., the parts without the log terms diverge at the null
cone
trough the origin.
\section{Discussion} \label{rsdisc}
In this article we have studied space-times which are asymptotically flat
at spacelike infinity, $i^0$. We have chosen to follow the ideas of
\cite{asht1}.
The differences are that we have used a weaker differentiable structure at
$i^0$
and put more conditions on the limits along $\spp$ and $\sm$.

The conditions on $\spp$ and $\sm$ are motivated by peeling property,
\cite{pen2},
and by the behaviour of electromagnetic fields in flat space-time. These
conditions eliminate (in the case of electromagnetic fields) some of the fields
that are ``obviously non-physical'' as a whole, but which still have the
expected fall off in spacelike directions at $i^0$.

The differentiable structure, the $C^{1^+}$ structure, allows, at least
potentially, the Weyl tensor to diverge in an unwanted way. If $r$ is a
parameter
satisfying (\ref{rhofrihet}), we know that $r{R_{abc}}^d$ is a regular
direction dependent tensor if we use a $C^{>1}$ differentiable structure at
$i^0$.
If we use only a $C^{1^+}$ structure, $r{R_{abc}}^d$ may diverge like $\log r$.
We have shown that in the latter case, the relevant part of $r{R_{abc}}^d$,
the ($r$ times the) Weyl tensor, $r{C_{abc}}^d$, is nevertheless regular
direction dependent. To conclude this, the limits along both $\spp$ and $\sm$
must be considered. This may be related to the discussion in \cite{ashtpen}.

In the electromagnetic case, the situation is similar. If we consider fields of
the form (\ref{emlog}), we can eliminate the logarithmic part only if we
include limits of the field along both future directed and passed directed
null directions. For instance, we can combine (\ref{laddskalimink}) and
(\ref{logexempelem}) to get a field which is smooth on the future null cone
through the origin.

An interesting question is of course if there exists a completion of the
Schwarz\-schild space-time where the metric is $C^{>0}$ in both spacelike and
null directions. If so, the antipodal property of theorem \ref{antipod}
will follow (by the same argument as used here).

Another interesting question is the relation to the decomposition of the
Weyl tensor into the electric and magnetic parts, \cite{asht1}. As
discussed in \cite{asht1}, the vanishing of
$B_{ab}$, the magnetic part of the Weyl tensor, is a necessary condition in
order to be able to define angular momentum at $i^0$. Using the tetrad of
section \ref{rtetrad}, one finds that $B_{ab}M^a\overline M^b$ corresponds
to the imaginary (and $E_{ab}M^a\overline M^b$ to the real) part of
$\gamma_2$ in the expansion (\ref{gamma2expan}). If the imaginary part of
$\gamma_2$ is zero for null directions, the same will hold for all directions.
Since $\gamma_2$ on e.g. $\spp$ is related to $\Psi_2$, there may be a
connection to the positive mass theorems.

We also hope that the use of $C^{1^+}$ differential structures which in a sense
allow logarithmic divergences at $i^0$, together with the
symmetry property of the rescaled Weyl tensor can shed some light on the so
called
logarithmic ambiguities, \cite{beigschmidt}, \cite{ashtlogamb} in the
completions
near $i^0$.
\section{Appendix A, A tetrad near $i^0$.}
\label{rtetrad}
In this appendix we will justify our choice of the direction dependent tetrad
in section \ref{ubianchi}.

Choose any future directed, timelike vector $v^a \in T_{i^0}M$, such that
\newline
$v^av_a=~2$. We then take a chart with coordinates $x^1=t$,
$x^2=x$, $x^3=y$ ,$x^4=z$
containing~$i^0$. By a translation and a linear transformation we may
assume that
$x^i(i^0)=0$ and that the metric at $i^0$ has components diag$(1,-1,-1,-1)$.
We can also arrange so that $v^a=\sqrt 2 \pp t$ at $i^0$.

Let us also define $\rho$ via $\rho^2=t^2+x^2+y^2+z^2$ and $r$ via
$r^2=x^2+y^2+z^2$. Recall that the metric $g_{ab}$
has components $g_{ij}$ that can be written
\beqn
g_{ij}=\eta_{ij}+\rho\!\log\! \rho\; k_{ij}
\eeqn
where each $k_{ij}$ is constant along rays and the $C^{>0}$-functions
$\eta_{ij}$ have the Min\-kowski metric as limit at $i^0$.

We will now define the direction dependent null vectors $N^a$ and $L^a$. It
will be convenient to
adopt the convention that $h$ stands for any $C^{>0}$-function which is zero at
$i^0$, and that $k$ stands for any function which is constant on rays. We then
define $N^a$ as
\beqn
N^a=\frac{1}{\sqrt 2}[(\pp t)^a+(1+\alpha)(\pp r)^a],\quad(\pp r)^a=
(\frac{x}{r}\pp x+\frac{y}{r}\pp y+\frac{z}{r}\pp z)^a
\eeqn
where $\alpha$ is chosen such that we get a null vector.
Using our convention, we have $(\pp t)^a(\pp t)_a=1+h+\rho \log\!\rho k$,
$(\pp t)^a(\frac{x}{r}\pp x+\frac{y}{r}\pp y+\frac{z}{r}\pp z)_a=h+\rho
\log\!\rho k$
and $(\frac{x}{r}\pp x+\frac{y}{r}\pp y+\frac{z}{r}\pp z)^a
(\frac{x}{r}\pp x+\frac{y}{r}\pp y+\frac{z}{r}\pp z)_a=-1+h+\rho \log\!\rho k$.
Thus, the condition that $N^a$ is a null vector implies that
\beqn
1+h+\rho \log\!\rho k+2(1+\alpha)(h+\rho \log\!\rho k)- \\
(1+\alpha)^2(1+h+\rho \log\!\rho k)=0.
\eeqn
This can be written
\beqn
\alpha^2(1+h+\rho\log\!\rho k)+2\alpha(1+h+\rho\log\!\rho
k)+h+\rho\log\!\rho k=0
\eeqn
from which we conclude that $\alpha$ is also of the form
\beqn
\alpha=h+\rho\log\!\rho k,
\eeqn
i.e., $\alpha$ is a $C^{0^+}$-function with $\alpha(i^0)=0$.

Similarly, we define $L^a$ as
\beqn
L^a=\frac{(1+\gamma)}{\sqrt 2}[(\pp t)^a-(1+\beta)(\pp r)^a],
%(\frac{x}{\rho}\pp x+\frac{y}{\rho}\pp y+\frac{z}{\rho}\pp z)^a,
\eeqn
where both $\beta$ and $\gamma$ are $C^{0^+}$-function that are zero at $i^0$,
and where we first choose $\beta$ so that $L^a$ becomes a null vector, and
then choose $\gamma$ so that $L^aN_a=1$.

To complete the tetrad, we must define a complex null vector $M^a$ such that
$M^aL_a=M^aN_a=0$, $M^aM_a=0$, $M^a\overline M_A=-1$.
We define the imaginary part of $M^a$ via
\beqn
\mbox{Im}(M_a)=\frac{1}{\sqrt 2}\frac{u_a}{\sqrt{-g^{ab}u_au_b}},\quad
u_a=\frac{y}{\sqrt{x^2+y^2}}(dx)_a-\frac{x}{\sqrt{x^2+y^2}}(dy)_a,
\eeqn
and the real part of $M^a$ via
\beqn
\mbox{Re}(M_a)=\frac{1}{\sqrt 2}\frac{w_a}{\sqrt{-g^{ab}w_aw_b}},\quad
w_a=dx_{[a}dy_b dz_{c]}(\pp r)^b u^c.
\eeqn
We now have
\beqn
L^av_a=1,\quad N^av_a=1,\quad  L^a+N^a=v^a
\eeqn
at $i^0$.

In order to  derive e.g. (\ref{spinder2}), we have to calculate
\beqn
\Dp o^A=\lim_{i^0}r\D^{\prime o} o^A=\lim_{i^0}-r\rho\iota^A,
\eeqn
etc.\ where we have used that $\D^{\prime o} o^A=-\rho\iota^A$, \cite{pen1}.
Thus we must calculate $\lim_{i^0}-r\rho$, where the spin coefficient $\rho$ is
defined as $\rho=m^a\overline m^b\nabla_b l_a$. We have
\beqn
\rho&\!\!\!\!=&\!\!\!\!m^a\overline m^b\nabla_b l_a=m^a\overline
m^b\nabla_b (eL_a) \nonumber \\
&\!\!\!\!=&\!\!\!\! em^a\overline m^b\nabla_b L_a+L_am^a\overline
m^b\nabla_b e=
e m^a\overline m^b\nabla_b L_a.
\eeqn
Writing $L_a=\widetilde L_a+h+\rho \log \!\rho \,k$,
$M_a=\widetilde M_a+h+\rho \log \!\rho \,k$ and $M_a=\widetilde M_a+h+\rho
\log \!\rho \,k$, where
\beqn
\!\widetilde N^a\!&\!\!\!\!=&\!\!\!\!\!\frac{1}{\sqrt 2}(\pp t + \pp
\rho)^a, \quad
\!\widetilde L^a\!=\!\frac{1}{\sqrt 2}(\pp t - \pp \rho)^a, \nonumber \\
\!\widetilde M^a\!&\!\!\!\!=&\!\!\!\!\!\frac{1}{\sqrt 2 \rho}(\pp \theta -
\frac{i}{\sin \theta}\pp \varphi)^a,
\eeqn
and using that $r{\Gamma^i}_{jk}=
h+\rho \log \!\rho \,k$, we find that
\beqn
r\rho=e(r\widetilde M^a\overline{\widetilde M^b}\widetilde \partial_b
\widetilde L_a
+h+\rho \log \!\rho \,k),
\eeqn
where $\widetilde \partial_a$ is the coordinate derivative. This means that
we can use the
results from flat space-time, i.e., use that
$r\widetilde M^a\overline{\widetilde M^b}\widetilde \partial_b \widetilde
L_a=1$ so that
\beqn
\lim_{i^0}r\rho=e,\;\mbox{i.e.}\;\Dp o^A=-e\iota^A
\eeqn
at $i^0$.

The commutator relations are derived in a similar manner, for instance
\beqn
\Ph\D-\D\Ph&\!\!\!\!=&\!\!\!\!\lim_{i^0}\{r\Ph^o(r\D^o)-r\D^o(r\Ph^o)\}=\lim_{i^
0}-er\D^o+
\lim_{i^0}r^2\{\Ph^o\D^o-\D^o\Ph^o\} \nonumber \\
&\!\!\!\!=&\!\!\!\!-e\D+
\lim_{i^0}r^2\{\overline \rho \D^o+\sigma \D^{\prime o}-\overline \tau '
\Ph^o-\kappa \Ph^{\prime o}-p(\rho ' \kappa -\tau ' \sigma+ \Psi_1)
\nonumber \\
&&\!\!\!\!-q(\overline \sigma ' \overline \kappa '-\overline
\rho\hspace{0.1mm} \overline \tau '
+\Phi_{01})\}=-e\D+e\D=0.
\eeqn
Here we have used that $\lim_{i^0}r^2\Psi_1=\lim_{i^0}r^2\Phi_{01}=0$, which
follows directly from \newline $\lim_{i^0}r^2R_{abcd}=0$. In the same way,
we find
the commutator relations
\beqn
\Ph\Php-\Php\Ph&\!\!\!\!=&\!\!\!\!-e\Php-\ei\Ph,\\
\Ph\D-\D\Ph&\!\!\!\!=&\!\!\!\!0, \\
\D\Dp-\Dp\D&\!\!\!\!=&\!\!\!\!(p-q),
\eeqn
plus the primed and conjugated versions.

\end{document}